\documentclass[12pt]{article}
\pdfoutput=1
\usepackage[utf8]{inputenc}
\usepackage{epsf}
\usepackage{colortbl}
\usepackage{amsmath}
\usepackage{amsfonts}
\usepackage{amssymb}
\usepackage{graphicx}
\usepackage{color}
\usepackage{psfrag}
\usepackage{cite}
\usepackage{hyperref}
\usepackage{tikz}
\usepackage{ifpdf}

\newcommand{\bmat}{\left(\begin{array}}
\newcommand{\emat}{\end{array}\right)}

\def\yzero{\smash{\hbox{$y\kern-4pt\raise1pt\hbox{${}^\circ$}$}}}

\def\beq{\begin{equation}}
\def\eeq{\end{equation}}
\def\beqa{\begin{eqnarray}}
\def\eeqa{\end{eqnarray}}

\def\-{\hphantom{-}}

\def\s2{\frac{1}{\sqrt2}}

\def\beq{\begin{equation}}
\def\eeq{\end{equation}}
\def\beqa{\begin{eqnarray}}
\def\eeqa{\end{eqnarray}}

\def\IF{\relax{\rm I\kern-.18em F}}
\def\II{\relax{\rm I\kern-.18em I}}

\def\Dsl{\,\raise.15ex\hbox{/}\mkern-13.5mu D} %this one can be subscripted
\def\G{\Gamma}

\newcommand{\eq}[1]{(\ref{#1})}

\catcode`\@=11   
\newdimen\@rotdimen
\newbox\@rotbox  

\def\@vspec#1{\special{ps:#1}}%  passes #1 verbatim to the output
\def\@rotstart#1{\@vspec{gsave currentpoint currentpoint translate
   #1 neg exch neg exch translate}}% #1 can be any origin-fixing transformation
\def\@rotfinish{\@vspec{currentpoint grestore moveto}}% gets back in synch 
\def\@rotr#1{\@rotdimen=\ht#1\advance\@rotdimen by\dp#1%
   \hbox to\@rotdimen{\hskip\ht#1\vbox to\wd#1{\@rotstart{90 rotate}%
   \box#1\vss}\hss}\@rotfinish}
\def\@rotl#1{\@rotdimen=\ht#1\advance\@rotdimen by\dp#1%
   \hbox to\@rotdimen{\vbox to\wd#1{\vskip\wd#1\@rotstart{270 rotate}%
   \box#1\vss}\hss}\@rotfinish}%
\def\@rotu#1{\@rotdimen=\ht#1\advance\@rotdimen by\dp#1%
   \hbox to\wd#1{\hskip\wd#1\vbox to\@rotdimen{\vskip\@rotdimen
   \@rotstart{-1 dup scale}\box#1\vss}\hss}\@rotfinish}%
\def\@rotf#1{\hbox to\wd#1{\hskip\wd#1\@rotstart{-1 1 scale}%
   \box#1\hss}\@rotfinish}%
\def\rotate{\@ifnextchar[{\@rotate}{\@rotate[l]}}
\def\@rotate[#1]#2{\setbox\@rotbox=\hbox{#2}\@nameuse{@rot#1}\@rotbox}

\catcode`\@=12
\topmargin
-1.5cm
\textwidth
15.5cm
\textheight
23.5cm
\oddsidemargin
0.7cm
\evensidemargin
1.2cm
\begin{document}

%----------------------------------------------------------------------%
%  numbering equations with section number
%----------------------------------------------------------------------%
\makeatletter
\@addtoreset{equation}{section}
\makeatother
\renewcommand{\theequation}{\thesection.\arabic{equation}}
%----------------------------------------------------------------------%
%  title page
%----------------------------------------------------------------------%
\hypersetup{pageanchor=false}
\pagestyle{empty}
%\vspace*{1.0in}
\rightline{ }

\vspace{3cm}

\begin{center}
\LARGE{\bf Are tiny gauge couplings out of the Swampland? \\[12mm] }
\large{M. Montero$^{1}$\\[4mm]}
\footnotesize{
${}^1$Institute for Theoretical Physics and
Center for Extreme Matter and Emergent Phenomena,\\
Utrecht University, Princetonplein 5, 3584 CC Utrecht, The Netherlands
}

\vspace*{5mm}

\small{\bf Abstract} \\%[5mm]
\end{center}
\begin{center}
\begin{minipage}[h]{\textwidth}
There is significant evidence suggesting that continuous global symmetries are always gauged in quantum gravity. However, very weakly gauged symmetries seem global to an effective field theory expansion in powers of Newton's constant. We show that, at least for Einsteinian quantum gravity on AdS, such extremely weak gaugings are indeed in the Swampland: Consistency with AdS black hole thermodynamics requires the bulk gauge coupling $g^2$ not to vanish faster than $\sim\exp(\ell^{d-1}/G)$, where $\ell$ is the $AdS_{d+1}$ radius and $G$ is Newton's constant as we take the $G\rightarrow0$ limit. This translates to a constraint in the dual large $N$ CFT, namely, that the two-point function coefficient of the current $C_J$ cannot grow faster than $\exp(N^2)$ in the large $N$ limit.  We also recover a previously known logarithmic relationship between the cutoff of the effective field theory in AdS, Planck's mass, and the AdS radius.
\end{minipage}
\end{center}
\newpage
\hypersetup{pageanchor=true}
%----------------------------------------------------------------------%
%  Resetting of counters
%----------------------------------------------------------------------%
\setcounter{page}{1}
\pagestyle{plain}
\renewcommand{\thefootnote}{\arabic{footnote}}
\setcounter{footnote}{0}
%----------------------------------------------------------------------%
%  Paper begins
%----------------------------------------------------------------------%

\tableofcontents 

\vspace*{1cm}

\section{Introduction}
Recently there has been a renewed interest in the Swampland \cite{Vafa:2005ui,Ooguri:2006in,Adams:2006sv}, the idea that not every effective field theory one can write down is actually consistent with quantum gravity. This opens up the possibility of exploring generic features that consistent theories must satisfy, rather than focusing on specific models. The most famous example of such a constraint is the Weak Gravity Conjecture \cite{ArkaniHamed:2006dz}, which has been applied extensively to try and constrain models of large field inflation or relaxation \cite{delaFuente:2014aca,Rudelius:2014wla,Rudelius:2015xta,Montero:2015ofa,Brown:2015iha,Bachlechner:2015qja,Hebecker:2015rya,Brown:2015lia,Junghans:2015hba,Palti:2015xra,Heidenreich:2015nta,Kooner:2015rza,Heidenreich:2015wga,Ibanez:2015fcv,Montero:2016tif,Heidenreich:2016aqi,Hebecker:2016dsw,Saraswat:2016eaz,Herraez:2016dxn,Ooguri:2016pdq,Cottrell:2016bty,Hebecker:2017wsu}. It has also been extended to situations with scalar fields \cite{Palti:2017elp} or axionic black holes \cite{Hebecker:2017uix}, refined to a sublattice version \cite{Heidenreich:2016aqi,Montero:2016tif}, and explored in the AdS/CFT context \cite{Nakayama:2015hga,Montero:2016tif}. Other recent examples of proposed  Swampland criteria are the Refined Swampland Conjecture of \cite{Klaewer:2016kiy} or the  Chern-Simons pandemic \cite{Montero:2017yja}. 

Arguably, the first example of a Swampland-like criterion was the proposed absence of continuous global symmetries in quantum gravity \cite{Abbott:1989jw,Coleman:1989zu,Kallosh:1995hi,Banks:2010zn}. It is also the most clearly established so far, with a convincing argument coming from black hole heuristics \cite{Susskind:1995da}, worldsheet proofs in the context of string theory \cite{Banks:1988yz}, and a beautiful identification with Noether's theorem in the AdS/CFT correspondence \cite{Beem:2014zpa}. 

If a strictly global symmetry is forbidden in quantum gravity, the next natural question is to think about very weakly gauged ones. In fact, this line of reasoning leads to the heuristic justification for the WGC provided in \cite{ArkaniHamed:2006dz}. More recently, \cite{Saraswat:2016eaz} has advocated a weaker bound, relating Planck's mass, the gauge coupling, and the cutoff of the effective field theory, based solely on entropy bounds. This note explores the consequences of essentially the same argument, but in the AdS/CFT context. Here, one can immediately see that there is a problem with a tiny coupling: As reviewed in the next Section, the AdS effective field theory is actually defined via a $1/N$ expansion of the dual CFT. If the gauge coupling (defined via the current two-point function) goes as $\exp(-N^2)$, it will be invisible to perturbation theory; from the point of view of the effective field theory, the symmetry remains effectively ungauged.  

In the bulk theory, one expects nonperturbative quantum-gravitational effects such as virtual Planck-sized black holes to contribute to every correlator or vertex in the quantum effective action. These effects are expected to be of order $\exp(-\ell^{d-1}/G)\sim \exp(-N^2)$. From a bulk perspective, both these nonperturbative effects as well as ordinary effective field theory computations contribute to the the CFT two-point function. If the two-point function itself goes as $\exp(N^2)$, one can conceive a scenario where $g=0$ and the theory is still compatible with the CFT result, due entirely to the gravitational contributions.

 In other words, the CFT Noether's theorem is technically compatible with a global symmetry in the bulk effective theory, since the symmetry can be very weakly gauged by gravitational effects.  These tiny or even vanishing gauge couplings in the EFT seem completely harmless, but at the same time never show up in known stringy examples, so one may wonder if they are in the Swampland or not. The main point of this note is that they indeed are, at least for AdS theories with an Einstein holographic dual, since they lead to conflict with generic features of black hole thermodynamics. 

This result translates in CFT language to a bound on the behavior of the CFT current two-point function coefficient as a function of $N$. The bootstrap program (see \cite{Simmons-Duffin:2016gjk} for a recent review) has been remarkably successful in recent years in constraining CFT data, both numerically and analytically, also in the context of large $N$ gauge theories (see e.g. \cite{Heemskerk:2009pn,Fitzpatrick:2011ia,Fitzpatrick:2012yx} among many others). The bound derived in this note would be difficult to obtain using bootstrap methods, as these are naturally tied to the $1/N$ expansion in holographic theories.

We also recover a version of the logarithmic relationship between the effective field theory cutoff, Planck's mass, and gauge coupling derived in flat space in \cite{Saraswat:2016eaz}; the bound is weaker than its flat space counterpart but the tradeoff is that it now becomes a sharp statement that can be proven rigorously, given the assumptions.

The note is organized as follows: Section \ref{adslore} reviews the AdS/CFT lore relevant to the problem, and also reformulates in CFT language the statement that a gauge symmetry with a tiny gauge coupling becomes invisible to the effective field theory.  Section \ref{sec:bht} provides a brief review of AdS black hole thermodynamics and introduces the relevant ensembles and their partition functions. Section \ref{bounds} contains the main result, deriving two bounds on the large $N$ behavior of the gauge coupling and its relationship with the effective field theory cutoff (a similar bound was derived in \cite{Saraswat:2016eaz}); finally, Section \ref{conclus} presents a discussion of the results. Some details have been relegated to an Appendix.

\section{Review of some aspects of holographic large \texorpdfstring{$N$}{N} gauge theories}\label{adslore}
We will start by reviewing some of the basic aspects of a large $N$ gauge theory with an Einstein holographic dual. The spirit and most details of the discussion are drawn from \cite{ElShowk:2011ag}. 

\subsection{Holography and the large \texorpdfstring{$N$}{N} limit}

The gist of the AdS/CFT correspondence is that the correlators of a (not necessarily supersymmetric\footnote{Although, according to the conjecture in \cite{Ooguri:2016pdq}, maybe nonsupersymmetric examples do not exist!}) quantum theory of gravity in global $AdS_{d+1}$ define a $d$-dimensional local\footnote{Meaning it has a stress-energy tensor.} CFT. The spectrum of local operators of the CFT is in one-to-one correspondence with states in the AdS theory, and their dimensions correspond to the energies of the bulk states. 

Since, given a quantum theory of gravity in AdS, we seem to get a CFT, we might try to reverse the logic, asking the question of which CFT's actually correspond to a bulk AdS theory with a weakly coupled Einstein gravity dual. There are some features we can learn immediately from the spectrum that AdS EFT's have: There are a few operators, of spin $s\leq 2$, which describe the (small number of) light fields propagating in AdS, which have a low dimension. There is a parametric gap in energies after which we reach black hole microstates, and the degeneracy increases exponentially.

The set of operators dual to small deviations from the AdS background has additional structure. Namely, we must reproduce the approximate Fock space structure that almost free theories display. This means that if one has two operators $O_1,O_2$, there should be a third one $O_{1+2}$, with dimension $\Delta_{1+2}\approx \Delta_1+\Delta_2$. 

The prime example of a theory displaying all of the above properties is a large $N$ gauge theory. In these theories one usually has a few light mesonic operators (the single-trace operators), separated by a gap (which grows parametrically with $N$) from a huge number of states in the deconfined phase. 

Large $N$ perturbation theory enjoys a number of properties which are mapped to properties of the quantum gravity theory in AdS. Correlators of single-trace operators factorize modulo $1/N$ corrections,
\begin{align}\langle OO\rangle=1,\quad \langle OOO\rangle\sim1/N,\quad \langle OOOO\rangle\sim \langle OO\rangle\langle OO\rangle+\mathcal{O}\left(\frac{1}{N^2}\right),\label{1Nexp}\end{align}
 so that one can use Wick's theorem to compute an arbitrary $n$-point function in a large $N$ expansion in which connected $n$-point functions scale as $N^{-n/2}$. To match this to a bulk description, it is easiest to notice that in large $N$ gauge theories, the two-point function of the stress-energy tensor $C_T$, is proportional to $N^{d/2}$. On the other hand, computing the same two-point function in Einstein's theory in AdS yields the relation \cite{Liu:1998bu,Mueck:1998ug,Polishchuk:1999nh}\footnote{There is a missing factor of 4 in eq. (43) of \cite{Mueck:1998ug}. This is necessary for agreement with the results of \cite{Barnes:2005bw,Barnes:2005bm,Nakayama:2015hga}, as well as for reproducing the Brown-Henneaux formula $c=3\ell/2G$ in $d=2$.}
\begin{align}C_T=\frac{d(d+1)}{(d-1)}\frac{\Gamma(d)}{\pi^{d/2} \Gamma(d/2)} \frac{\ell^{d-1}}{\kappa_{d+1}^2},\quad \kappa_{d+1}^2=8\pi G.\label{CT}\end{align}
This means that the large $N$ expansion of the gauge theory is secretly a loop expansion in powers of Newton's constant in the bulk. Having large $N$, and thus large $C_T$, ensures that gravity is weakly coupled in the bulk and that the perturbative approach is appropriate.

In the rest of this note, we will abstract these nice properties of gauge theory to a generic CFT dual to Einsteinian gravity in the bulk. Namely, we will consider families of CFT's indexed by a parameter $N$ (which may not be an integer), with the property that $C_T$ grows as $N^{2}$, and that in the large $N$ limit there is a special set of operators of low dimension and spin $\leq2$, which does not grow parametrically with $N$, and such that their correlators factorize. From the bulk perspective, correlators of these operators may be computed order by order in $1/N$ as a Witten diagram loop expansion in powers of Newton's constant. Notice that the $C_T$ scaling with $N$ we have chosen is characteristic of four-dimensional gauge theories, but for concreteness we will be applying it for any $d$. This is merely a bookkeeping device; every result can be recast in terms of invariant quantities like $C_T$. 

\subsection{Symmetries and gauge fields}
We will need some more details of the correspondence for the particular case of a conserved current/gauge field. Suppose the CFT has a continuous global $U(1)$ symmetry. Because the correspondence is an exact identification between theories, the bulk theory has a $U(1)$ symmetry as well.

 As discussed in the Introduction, there are several arguments pointing to the conclusion that quantum gravity does not like global symmetries (this is clearly the case for continuous symmetries; but every concrete discrete example seems to be gauged, as well). In fact, in AdS/CFT it is possible to make a very strong case for this. Suppose the CFT admits a Lagrangian description. Then, Noether's theorem guarantees the existence of a conserved current which generates the symmetry.  In a CFT, the statement that the current is conserved becomes very strong - it fixes the dimension to be $d-1$ and, if the current is in addition a primary field\footnote{Conserved descendant currents can arise e.g. as exterior derivatives of a non-conserved $2$-form primary $B$, such that $d*B=*J$. In this case, charged operators under $J$ must have strings charged under $B$ attached to them.}, the two-point function is fixed to
 \begin{align}\langle J_\mu (x) J_\nu(0)\rangle=C_J\frac{I_{\mu\nu}(x)}{x^{2d-2}},\quad I_{\mu\nu}(x)\equiv \delta_{\mu\nu}-2\frac{x_\mu x_\nu}{x^2}.\end{align}
 The two-point function coefficient, $C_J$, can be expanded in the $1/N$ expansion. One can then compute the generating function of correlators of $J$,
 \begin{align}Z[A]=\left\langle\exp\left(\int J_\mu A^\mu\right)\right\rangle,\end{align}
and, according to standard AdS/CFT lore, $A_\mu$ is the boundary value of a bulk gauge field. From the AdS side, the leading contribution to $C_J$ comes from the Maxwell term in the action, and relates $C_J$ to the bulk gauge coupling $g$ and the AdS radius $\ell$ as \cite{Freedman:1998tz}
\begin{align}C_J=\frac{d-2}{2}\frac{\Gamma(d)}{\pi^{d/2}\Gamma(d/2)}\frac{\ell^{d-3}}{g^2}.\end{align}
Typically, the current $J$ is a single-trace operator, and according to \eq{1Nexp}, the three-point function $\langle \mathcal{O}_q J \bar{\mathcal{O}}_q\rangle $ is of order $1/N$, if we normalize $\langle JJ\rangle\sim 1$, $\langle \mathcal{O}_q  \bar{\mathcal{O}}_q\rangle\sim1$. From the bulk perspective, the leading contribution to this three-point function is proportional to $g$, which leads us to conclude $g\sim 1/N$, and hence $C_J\propto N^2$, as is the case in known four-dimensional superconformal theories \cite{Barnes:2005bw,Nakayama:2015hga}. 

The above is the standard behavior of $C_J$. However, it is perfectly consistent to imagine holographic theories in which $C_J$ scales with a different power of $N$. Admittedly, in this case $J$ would not be a single-trace operator in the way defined in \eq{1Nexp}, but there is no a priori reason why the three-point functions of all the light operators (those whose dimension does not depend on $N$ in the large $N$ limit) should all scale with the same power of $N$.

If $C_J$ grows as $N^{\alpha}$, with $\alpha>1$, the bulk gauge coupling vanishes faster than it does in ordinary large $N$ gauge theories with a holographic dual, but there seems to be nothing wrong with this: The $1/N$ expansion, and its bulk description in terms of Witten diagrams, remains essentially the same. 

Things are different if $C_J$ grows really fast, as $e^{N^2}$. Since the bulk effective field theory arises as a $1/N$ expansion, the gauge coupling is effectively zero in perturbation theory. Any $n$-point functions involving currents will vanish in perturbation theory. For all intends and purposes, the effective field theory has a global symmetry, even though the CFT satisfies Noether's theorem. There is a bulk gauge field, but it couples only gravitationally.

Notice that the quantum-corrected gauge coupling (the coefficient of the Maxwell term in the quantum effective action), while tiny, is  nonzero, so technically the symmetry is gauged, even if this cannot be seen in the $1/N$ expansion. As discussed in the Introduction, its value is set by some nonperturbative gravitational effect, suppressed as $e^{N^2}\sim\exp(-l^{d-1}/G_N)$. Maybe this is all one needs to be consistent with quantum gravity. The point is that the effective low-energy bulk lagrangian, which does not include quantum gravity effects, can have an arbitrarily small -even vanishing- gauge coupling, without contradicting the CFT lore just discussed.

Because of this, one could think that maybe tiny gauge couplings are just fine. The rest of this note is devoted to showing that this is not the case, at least for CFT's with an Einstein holographic dual. 

\section{Relevant aspects of black hole thermodynamics}\label{sec:bht}
We will now review the essentials of AdS black holes relevant to our analysis, following closely  \cite{Chamblin:1999tk,Chamblin:1999hg}.

The action of the corresponding $AdS_{d+1}$ Einstein-Maxwell sector is
\begin{align}\int d^{d+1}x\left(\frac{R}{2\kappa_{d+1}^2}+\frac{d(d-1)}{\ell^2}\right)-\frac{1}{4g^2} F_{\mu\nu}F^{\mu\nu}.\label{emac}\end{align}
Newton's constant is given by $\kappa_{d+1}^2=8\pi G$. We normalize electric charges to be quantized, so that \cite{Heidenreich:2015nta}
\begin{align}Q=\frac{1}{g^2}\int *F\end{align}
is an integer. We will also assume that \eq{emac} is a consistent truncation of the full theory. This happens in known examples in string theory \cite{Chamblin:1999tk,Bhattacharyya:2010yg}; from a practical point of view, this means that classical solutions of the equations of motion coming from the action \eq{emac} will also be solutions of the full theory.   In particular, the theory \eq{emac} admits charged black hole solutions (which we will call Reissner-Nordstrom-AdS). The metric is 
\begin{align}ds^2=-V(r) dt^2+\frac{dr^2}{V(r)} +r^2d\Omega_{d-1},\quad V(r)\equiv1-\frac{m}{r^{d-2}}+\frac{q^2}{r^{2d-4}}+\frac{r^2}{\ell^2}.\label{bhmetric}\end{align}
There is also a gauge field background,
\begin{align}A=\Phi-\frac{1}{c}\frac{\sqrt{2}g}{\kappa_{d+1}}\frac{q}{r^{d-2}} dt, \quad c\equiv\sqrt{\frac{2(d-2)}{d-1}}.\end{align}
$\Phi$ is a constant which has to be tuned to guarantee smoothness of the Euclidean solution, but which will be otherwise unimportant in the present context. In our normalization, the parameters $m,q$ are related to the ADM mass and quantized charge as (here $\omega_{d-1}$ is the volume of the unit $(d-1)$-dimensional sphere)
\begin{align}M=\frac{(d-1)\omega_{d-1}}{16\pi G} m,\quad Q=\sqrt{(d-2)(d-1)}\frac{\omega_{d-1}}{\sqrt{8\pi G} g} q.\label{normaliz}\end{align} 
For given $m,q$, there are two horizons $r_\pm$, given by the solutions to $V(r_\pm)=0$. The semiclassical entropy is given by the area law
\begin{align}S=\frac{A}{4G}=\frac{\omega_{d-1}}{4G}r_+^{d-1}\label{bhent}\end{align}
and the temperature $\beta$ is related to the horizon radius $r_+$ and $q$ as
\begin{align}q^2\ell^2=\frac{d}{d-2}r_+^{2d-2}+\ell^2r_+^{2d-4}-\frac{4\pi \ell^2r_+^{2d-3}}{(d-2)\beta}.\label{qfr}\end{align}

These black hole solutions have a prominent role in the thermodynamics of the theory. Consider the partition function of the CFT in a sector of fixed charge, $Z_{Q}(\beta)$. This is dual to the same $AdS$ theory with an extra boundary term and different boundary conditions to the gauge field, as reviewed in \cite{Marolf:2006nd} (see also \cite{Chamblin:1999tk,Chamblin:1999hg}). Alternatively, it may also be obtained as the Legendre transform of the partition function with a chemical potential $Z(\mu,\beta)$. In the semiclassical approximation, $Z_{Q}(\beta)$ is obtained as a sum over the saddles of the Euclidean path integral with the right asymptotic behavior. These include empty AdS with some charged matter fields as well as the  Euclidean continuation of \eq{bhmetric}. The action of this Euclidean section represents the contribution of the charged black hole microstates. The partition function satisfies
\begin{align}Z_{Q}(\beta)=\ldots+Z_{Q,BH}(\beta),\label{qwert}\end{align}
and in the large $N$ limit, where only the saddle with lowest free energy contributes \cite{Witten:1998qj}, we would actually get $Z_{Q}(\beta)=Z_{Q,BH}(\beta)$, for high enough temperatures.  The result is
\begin{align}\log Z_{Q,BH}(\beta)=\frac{\omega_{d-1}}{16\pi G \ell^2}\beta\left[r_+^d-\ell^2r_+^{d-2}-(2d-3)\frac{\ell^2q^2}{r_+^{d-2}}\right],\label{frebh}\end{align}
where $q$ is related to the quantized charge $Q$ via \eq{normaliz}. This can be obtained simply as $\exp(S-\beta M)$, where $S,M$ are given in \eq{normaliz}-\eq{bhent}, or directly as the Euclidean action using the background substraction method \cite{Chamblin:1999hg}. The phase diagram of this system as well as the regions where the AdS or the RN-AdS saddles dominate, are discussed in \cite{Chamblin:1999tk,Chamblin:1999hg}.

This whole story assumes that the RN-AdS black holes are stable. Although the consistent truncation \eq{emac} is good enough to write down the RN-AdS solutions and compute their action, their stability cannot be studied in the consistent truncation. Extra fields not present in the truncation can introduce new directions in field space along which the solution is actually unstable. In fact, one does not expect extremal or nearly extremal black hole solutions to be exactly stable in a consistent quantum theory of gravity: As discussed in \cite{Nakayama:2015hga}, the Weak Gravity Conjecture requires the presence of light charged fields, which generically trigger a Gubser-Mitra instability for large black holes \cite{Gubser:2000ec}, or Gregory-Laflamme \cite{Gregory:1994bj} for small ones. The endpoint of the condensation is either a ten-dimensional black hole, localized in the compact dimensions, or a black hole with charged scalar hair. Reference \cite{Bhattacharyya:2010yg} considers a concrete embedding of \eq{bhmetric} in string theory, including the coupling to one WGC scalar, showing that the endpoint of the condensation process is generically either a charged black hole with scalar hair or a solitonic configuration for the charged field. 

In the Euclidean version of the theory, the instability means that the Euclidean RN-AdS is in general no longer a local minimum of the free energy. Tachyonic fluctuations destroy its contribution to $Z_{Q}(\beta)$, so it seems that cannot write down \eq{qwert}. However, this difficulty has a simple solution.  Since RN-AdS is not a minimum, and we know from the dual CFT that the partition function $Z_{Q}(\beta)$ is not divergent, there must be another saddle with a lower free energy than RN-AdS, which is the actual dominant contribution in the large $N$ limit (this would be the endpoint of the condensation process described above). We do not know what the actual contribution of this saddle $Z_{Q,S}$ to $Z_{Q}(\beta)$ is, but clearly it is larger than  $Z_{Q,BH}(\beta)$\footnote{The one-loop corrections to $Z$ cannot be negative, since $Z_{Q,S}$ is the leading contribution to a manifestly positive CFT partition function, $Z_Q(\beta)$.}. As a result, even when the RN-AdS black hole is unstable, we have
\begin{align}Z_{Q}(\beta)\geq Z_{Q,BH}(\beta),\label{jey}\end{align}
so that the semiclassical free energy of RN-AdS is still a lower bound to the path integral\footnote{Strictly speaking $Z_{Q,BH}(\beta)$ is not well-defined because of the unbounded fluctuations of the tachyonic modes around the semiclassical solution. The discussion and \eq{jey} are meant to hold only at the semiclassical level, ignoring the subleading contributions coming from fluctuations around $Z_{Q,S}$}. 

Physically, \eq{jey} results from the second law of thermodynamics. Because we have assumed \eq{emac} to be a consistent truncation, the black holes \eq{bhmetric} are legitimate --if unstable-- states of the bulk theory. Their entropy \eq{bhent} is of course given by the black hole area. When the instability sets on and they decay, neither their charge nor their mass can change, because $AdS$ is a box. Since this is a spontaneous process, $\Delta S\geq0$, which means that the free energy of the RN-AdS black hole must be higher than that of the saddle describing the endpoint of the condensation process or, since $Z=\exp(-\beta F)$, \eq{jey}.  

Equation \eq{jey} also has implications for the standard canonical partition function, $Z(\beta)$\footnote{This is also $Z(\beta,\mu=0)$, the grand canonical partition function at vanishing chemical potential.}. As usual, this is defined as a sum over every state on the theory, with weight $e^{-\beta E}$. Thus, it can be written as
\begin{align}Z(\beta)=\sum_{\text{states}} e^{-\beta E} = \sum_{Q} \left(\sum_{\text{states w. charge $Q$}}e^{-\beta E}\right)=\sum_{Q} Z_{Q}(\beta)=Z_0(\beta)+\sum_{Q\neq0} Z_{Q}(\beta).\label{zdec}\end{align}
We have merely rearranged the sum over states in $Z(\beta)$ in terms of sectors of different charge. Equation \eq{jey} provides a bound on each of the terms of this sum; however, this doesn't mean that the Euclidean saddle that dominates $Z_{Q}(\beta)$ is also a saddle in $Z(\beta)$, precisely because the charge is allowed to fluctuate in the latter ensemble. 

On the other hand, the behavior of $Z(\beta)$ as a function of $\beta$ in a theory with an Einstein holographic dual has been understood since the inception of the AdS/CFT correspondence \cite{Witten:1998qj}. Just like we did for the charged partition function, one can write $Z(\beta)$ as a path integral over $S^{d-1}\times S^1$, and one must sum over all the saddles of the path integral with the right asymptotic behavior. In a theory with an Einstein holographic dual, meaning that the low-energy effective field theory consists only of Einstein gravity plus a small number of fields of spin lower than two, there are only two such saddles: Empty AdS space, dual to the low-temperature phase of the CFT, and the Schwarzschild-AdS black hole, which is \eq{bhmetric} with $Q=0$, dual to the high-temperature phase of the CFT. The free energy difference between the two is obtained by setting $q=0$ in \eq{frebh}. There is a phase transition at the Hawking-Page temperature, at $r_+=\ell$,
 \begin{align}\beta_{HP}\equiv\frac{2\pi\ell}{d-1}.\label{bHP}\end{align}
 The free energy  (adjusted so that AdS has zero free energy) to leading order in $N$ is therefore \cite{Witten:1998zw}
 \begin{align}\log Z_{\text{leading}}(\beta)\approx
\begin{cases}0& \text{if}\ \beta>\beta_{HP} \\ F(x_+^0) &\text{if}\  \beta<\beta_{HP} \end{cases}\ +\mathcal{O}(\log N). \label{key0}\end{align}

The function $F(x)$ and the dimensionless parameter $x_+^0$ are  defined in the Appendix. Let us emphasize that this leading large $N$ behavior of $Z(\beta)$ is generic of any Einsteinian theory in the sense described above. Since we only have a small number of light fields in the effective field theory, their fluctuations around the semiclassical solution will amount to a subleading, logarithmic correction. In the large $N$ limit, as interactions between the low-energy fields switch off, the thermodynamic behavior is controlled by the gravitational part of the action. We could take it as part of the definition of what it means for an AdS theory to have an Einstein dual.  As we will now see, imposing this leads to a lower bound on the $U(1)$ gauge coupling. 

\section{Bounds involving the gauge coupling}\label{bounds}
In the decomposition \eq{zdec}, it is clear that both the Schwarzschild-AdS black hole lives in (and dominates, at high temperatures) the $Z_0(\beta)$ part of the partition function. This means that, at high temperatures and to leading order in $N$, we have
\begin{align}Z_0\geq Z_{\text{leading}}.\end{align}
Combining with \eq{jey}, we obtain
\begin{align}Z(\beta)\geq Z_{\text{leading}}+\mathcal{Z}(\beta),\quad \mathcal{Z}\equiv \sum_{Q\neq0} Z_{Q,BH}(\beta),\quad\Rightarrow\quad Z(\beta)\geq Z_{\text{leading}}\left(1+ \frac{\mathcal{Z}(\beta)}{Z_{\text{leading}}}\right).\label{fff}\end{align}
On the other hand, to correctly reproduce the black hole thermodynamics of the previous Section, and in particular \eq{key0}, we should demand that, in the large $N$ limit, $Z(\beta)$ asymptotes to by $Z_{\text{leading}}$. This means that the second factor in the last equality of \eq{fff} must be a subleading correction; it must grow slower than $Z_{\text{leading}}$ in the large $N$ limit. Since both $Z_{\text{leading}}$ and $\mathcal{Z}$ go as $\exp(a N^2)$ in the large $N$ limit for some $a$, this is equivalent to imposing
\begin{align}Z_{\text{leading}}(\beta)\geq \mathcal{Z}(\beta).\end{align}
This inequality is a constraint on the theory. Using \eq{frebh}, we can evaluate the right hand side explicitly, which is done in the Appendix. The final result is 
\begin{align}\log\mathcal{Z}(\beta)\approx
\begin{cases}F\left(\frac{1}{\Lambda\ell}\right)+ \log\left(\frac{\kappa_{d+1}}{g\ell}\right)& \text{if}\ \beta>\beta_{HP} \\ \\ F\left(x_+\right) + \log\left(\frac{\kappa_{d+1}}{g\ell}\right)&\text{if}\  \beta<\beta_{HP} \end{cases}\ +\mathcal{O}(\log N). \label{key}\end{align}
Here, $F(x)$ (defined precisely in the appendix) is $-\beta$ times the free energy of a charged black hole of temperature $\beta$ and radius $\ell x$. The radius $x_+$ is that of a charge one black hole, slightly above the Hawking-Page temperature, and is defined in the Appendix. In what follows, we will ignore the subleading logarithmic corrections\footnote{On top of being subleading, the contribution discussed in the Appendix is not really reliable, since we are ignoring logarithmic corrections to $\log Z_{BH,Q}(\beta)$ itself anyway. There is no hidden $\log g$ dependence in the logarithmic corrections we are dropping, because we are in the perturbative regime and the loop expansion of \eq{emac} in powers of $g$ is reliable. This means that the quantum effective action is a smooth function of $g$, thus forbidding any $\log g$ terms. }. The bound $Z(\beta)\geq \mathcal{Z}(\beta)$ is naturally discussed in two different regimes, below and above the Hawking-Page transition:\begin{itemize}
\item For $\beta>\beta_{HP}$, the free energy of a small, charged black hole $-F\left((\Lambda\ell)^{-1}\right)$ is positive. In order to have a sensible Einstein theory in the bulk, the free energy should be dominated by the $AdS$ saddle, which has zero free energy. This imposes
\begin{align}-F\left(\frac{1}{\Lambda\ell}\right)> \log\left(\frac{\kappa_{d+1}}{g\ell}\right),\end{align}
which is a constraint on the cutoff. $F\left((\Lambda\ell)^{-1}\right)$ is a linear decreasing function of $\beta$, so the strongest bound is obtained at $\beta=\beta_{HP}$. We get a lower bound to the cutoff of a theory with gauge coupling $g$,
\begin{align}\log\left(\frac{\kappa_{d+1}}{g\ell}\right)<\frac{\omega_{d-1}}{4G}(\Lambda\ell)\Lambda^{1-d},\quad\text{or equivalently}\quad\Lambda^{d-2}< \frac{2\pi\omega_{d-1} \ell}{\log\left(\frac{\kappa_{d+1}}{g\ell}\right)\kappa_{d+1}^2}.\label{cutoffbound}\end{align}
Notice that we could drop the logarithmic dependence on $\kappa_{d+1}$, since it is a subleading logarithmic correction.
Eq. \eq{cutoffbound} then becomes a logarithmic relation between the gauge coupling of the gauge theory, just like the one advocated in \cite{Saraswat:2016eaz} based on the Bekenstein entropy bound. In fact, the first expression of \eq{cutoffbound} states precisely that the degeneracy of charged black hole states, which is controlled by $g$, should not exceed the Bekenstein-Hawking entropy of a black hole of radius $(\ell\Lambda)^{\frac{1}{d-1}}\Lambda^{-1}$. There are, however, several outstanding differences between the AdS result \eq{cutoffbound} and the flat space one in  \cite{Saraswat:2016eaz}. First of all, \eq{cutoffbound} involves directly the AdS radius; as a result, it is a weaker bound than the flat space one in \cite{Saraswat:2016eaz}. Clearly, this happens because the analysis of the low-temperature regime gets cut off at $\beta=\beta_{HP}$. Here, the contribution from large AdS black holes, which has no flat space parallel, starts dominating the partition function; as a result, while small black holes with $\beta>\beta_{HP}$ do exist, the free energy does not see them. By contrast, in flat space, it is possible to analyze smaller black holes, up to temperatures of order $\beta\sim\Lambda^{-1}$; In this case, one obtains a relationship 
\begin{align}\Lambda^{d-1}\lesssim\frac{1}{\kappa_{d+1}^2\log\left(\frac{\kappa_{d+1}}{g\ell}\right)},\label{cflat}\end{align}
which is precisely what \cite{Saraswat:2016eaz} obtains. Eq. \eq{cutoffbound} contains an extra factor $\Lambda\ell\gg 1$ on the right-hand side. On the other hand, while the result \eq{cflat} relies on a holographic entropy bound \cite{Bousso:1999xy,Bousso:2002ju} for which there is yet no general proof, \eq{cutoffbound} is a rigorous statement about quantum theories of gravity in $AdS$.

In principle, it should be possible to derive \eq{cflat}  the AdS context by working in the microcanonical ensemble, where the small AdS black hole is the dominant saddle \cite{Hawking:1982dh}.  Free energies should be replaced by entropies, and the sum \eq{zdec} would be replaced by the requirement that the entropy of the charged sectors shouldn't exceed that of a neutral black hole; this is exactly the argument in \cite{Saraswat:2016eaz}.

\item In the high temperature regime $\beta<\beta_{HP}$ we have a new scenario which, unlike the previous case, doesn't have an obvious parallel in flat space. Here, the bound becomes (removing the logarithmic dependence on $\kappa_{d+1}$)
\begin{align} F(x_+^0)-F(x_+)\approx -F'(x_+^0)\delta x_+ > \log\left(\frac{\ell^{\frac{d-3}{2}}}{g}\right)\sim \log C_J\label{48}. \end{align}
After the pertinent substitutions, we get
\begin{align}-F'(x_+^0)\delta x_+ =\frac{g^2}{\ell^{d-3}} \frac{\beta}{2\ell (d-2)}\left(\frac{\sqrt{4 \pi ^2-(\beta/\ell)^2 (d-2) d}+2 \pi }{(\beta/\ell) d}\right)^{2-d},\label{ssqw}\end{align}
which goes to zero as $N\rightarrow\infty$, because $g\rightarrow0$ in this limit. This would seem to imply that $\log C_J\rightarrow0$ in the large $N$ limit, which certainly is not the case, but we must not forget that \eq{ssqw} is only valid to leading order, that is to $\mathcal{O}(N^2)$. As a result, from \eq{48} and \eq{ssqw} we can only conclude that $\log C_J$ cannot grow faster than $N^2$ in the large $N$ limit. In other words, $C_J$ must grow slower than$\exp(\alpha N^2)$ for any $\alpha$ in the large $N$ limit. This immediately excludes a gauge coupling which vanishes as $\exp(-N^2)$, as advertised.  In fact, the leading correction to the free energy $F(x)$ away from the semiclassical regime is usually logarithmic in $N$ - there are no $N^\alpha \log N$ terms for $\alpha>0$. This is because in a weakly coupled Einsteinian theory, together with a small number of light fields, one expects to be able to compute the first quantum corrections using a loop expansion, which yields a $\log N$ correction. In this case, one can make the stronger statement that the gauge coupling cannot vanish faster than some power of $1/N$. 
\end{itemize}

\section{Discussion and outlook}\label{conclus}
One of the motivations for this work was to see exactly what goes wrong when a gauge coupling becomes tiny in a controlled setup. This is the heuristic setup used in \cite{ArkaniHamed:2006dz} to argue for the WGC. In the AdS context, rather than an argument for the WGC, we obtain a bound on the large $N$ behavior of the gauge coupling. This is independent of whether or not the effective field theory satisfies the WGC, as discussed in Section \ref{sec:bht}. 

The result of the previous Section closes the loophole discussed in the Introduction: In principle, one could imagine a CFT with a large $N$ limit and which satisfies Noether's theorem, such that the effective field theory  arising from the $1/N$ expansion has a global symmetry to all orders in $1/N$. However, consistency with the known features of the thermodynamics of a field theory with a Einstein holographic dual immediately requires that the gauge coupling should be visible in the large $N$ perturbation theory expansion dual to the bulk low-energy effective field theory -meaning that, in the effective field theory around AdS, the symmetry must be gauged. More concretely, the gauge coupling cannot vanish, in the large $N$ limit, as $\exp(-N^2)$ or faster. 

The result doesn't immediately help an effective field theorist trying to constrain the value of $g$ in the AdS bulk. The constraint only refers to the behavior of the gauge coupling as a function of $N$, while in any concrete model $N$ and $g$ are fixed. It is only if we add the extra assumption that the bulk is far out into the large $N$ regime - where every coupling is controlled by its leading large $N$ behavior, and the hierarchies valid in the large $N$ limit are respected\footnote{For instance, if g$\sim N^{-\alpha}$, $N$ should be large enough that $e^{-N^2}< N^{-\alpha}$.} - that the constraint becomes predictive: $\vert \log g\vert $ should be much smaller than $C_T$, given by \eq{CT}. Even so, the constraint is not terribly helpful: For instance, if our universe had the same absolute value of $\Lambda$ but opposite sign, we would get a constraint $g\gtrsim \exp(-10^{120})$.  

A more interesting way to think about the bound is from the CFT side - as a constraint on the current two-point function of large $N$ gauge theories. This constraint would be very difficult to derive in a bootstrap approach, since the four-point functions derived via Witten diagrams in AdS satisfy crossing symmetry automatically, order by order in $1/N$. 

The constraint on the large $N$ behavior of the gauge coupling constitutes a simple example of a Swampland constraint  - with the caveats mentioned above - shown to hold in a controlled setting.

We have also recovered a weaker version of the logarithmic bound in \cite{Saraswat:2016eaz} relating the cutoff scale to the gauge coupling and the gravitational scale, $\Lambda^{d-2}\lesssim (-\log g \kappa^2_{d+1}\ell)^{-1}$. The bound we obtain is not as strong as in flat space due to technical reasons; probably, the flat space result can be recovered in AdS as well by using the microcanonical ensemble.

The bounds discussed in this note are all very weak. In all likelihood, there are far better constraints applying to theories with global symmetries, like the magnetic Weak Gravity Conjecture or the similar version coming from the (sub)Lattice WGC \cite{Heidenreich:2016aqi,Montero:2016tif}. In contrast to the bounds presented here, which are backed by concrete computations in a controlled setup, all the WGC variants remain conjectural. Nevertheless, it is interesting to compare the two. The sublattice WGC requires  the existence of charged fields with masses $m\lesssim Q g$, for $Q=k\cdot n, n=1,2,\ldots$. If $g$ is tiny, and $k$ not too large, there are too many light fields in the effective field theory, so it is not Einsteinian gravity: The WGC (crucially, with $k$ not too large) seems to be telling us that tiny gauge couplings are not possible anyway.  This is consistent with what is seen many stringy examples, where lowering $g$ while keeping the Planck scale constant forces $\alpha'\rightarrow\infty$. 

Although it leads to a strong constraint, the sublattice version of the WGC admits a loophole: the index of the sublattice $k$ can, in principle, be arbitrarily large, and with this the effects of the WGC on the low-energy physics are diluted or disappear altogether. This might be a moot point since, as of now, there are no known examples of theories with a parametrically large sublattice index, and the current stringy evidence gives no reason to suspect this to be possible. By looking instead at the general thermodynamic arguments presented in this note, we exchange reach for control - the AdS bound obtained here stands on more solid footing, since the derivation only involves generic features of AdS effective theories which are well under control.  

\vspace{0.5cm}

\subsection*{Acknowledgements}
It is a pleasure to thank J. L. F. Barb\'{o}n, Anton de la Fuente, Thomas Grimm, Luis Iba\~{n}ez, Gary Shiu, Pablo Soler, Irene Valenzuela, Gianluca Zoccarato and especially William Cottrell, Joao Gomes and Prashant Saraswat for very useful discussions and comments. I am supported by a  postdoctoral fellowship from ITF, Utrecht University.

%----------------------------------------------------

\appendix

\section{Charged black hole contribution to the partition function}
In this Appendix we derive equation \eq{key} of the main text. We need to evaluate 
\begin{align}\mathcal{Z}(\beta)\equiv \sum_{Q\neq0} Z_{Q,BH}(\beta)=\sum_{Q\neq0} \exp\left[\frac{\omega_{d-1}}{2\kappa_{d+1}^2 \ell^2}\beta\left(r_+^d-\ell^2r_+^{d-2}-(2d-3)\frac{\ell^2q^2}{r_+^{d-2}}\right)\right].\end{align}
For fixed $\beta$, the charge $q^2$ is given as a function of the radius $r_+$ by \eq{qfr}. This is not a monotonic function, which means that sometimes there are up to three different black hole solutions with the same charge; in this case, we take $Z_{Q,BH}$ to be the sum of all of them. In the large $N$ limit, the sum over $Q$ becomes an integral, which we rewrite in terms of $x\equiv r_+/\ell$. The integration region $R$ is determined by requiring $Q^2\geq1$ or, equivalently, 
\begin{align}q^2\geq q_0^2=\frac{g^2\kappa_{d+1}^2}{(d-2)(d-1)\omega_{d-1}^2}.\label{qft}\end{align}
Using \eq{qfr}, this becomes $R=[x_{\text{min}},x_-]\cup[x_+,\infty)$, with
\begin{align}x_\pm&=x_\pm^0+\delta x_\pm,\quad x^0_{\pm}=\frac{\ell \left(2 \pi \pm\sqrt{4 \pi ^2-(\beta/\ell) ^2 (d-2) d} \right)}{\beta  d},\nonumber\\\delta x_\pm&= \frac{(\beta/\ell) (x^0_\pm)^{5-2 d}}{2 (d-1) \left((\beta/\ell) \left((d-1) d (x^0_\pm)^2+(d-2)^2\right)+2 \pi  (3-2 d) x^0_\pm\right)}\frac{g^2 \kappa ^2}{\ell^{2d-4}\omega_{d-1}^2},\end{align}
to first order in $g^2\kappa^2/\ell^{2d-4}$. 
When the discriminant vanishes, the integration region becomes $x>0$. This happens in the low-temperature regime, close to extremality, when
\begin{align}\beta>\beta_c\equiv\frac{2\pi\ell}{\sqrt{d(d-2)}}.\end{align}
Thus we obtain
\begin{align}\mathcal{Z}(\beta)&\approx \frac{\ell^{d-2}\omega_{d-1}}{g\kappa_{d+1}}\int_{R} dx\, x^{d-3}\, \vert G(x)\vert\exp(F(x)),\nonumber\\F(x)&\equiv -\frac{\omega_{d-1}\ell^{d-1}}{16\pi G}\frac{2  x^{d-2} \left(2 \pi  (3-2 d)x+(\beta/\ell)  (d-1) \left((d-1) x^2+d-2\right)\right)}{(d-2) },\nonumber\\G(x)&\equiv\sqrt{d-1}\frac{(2 \pi  (3-2 d) x+(\beta/\ell)  \left(d(d-1)  x^2+(d-2)^2\right))}{\sqrt{(\beta/\ell)  \left((\beta/\ell)  \left(d x^2+d-2\right)-4 \pi  x\right)}}.\end{align}
We now need to specify $x_{\text{min}}$. Naively, \eq{qft} suggests $x_{\text{min}}=0$, which corresponds to very small charged black holes. However, although we assumed the Einstein-Maxwell action \eq{emac} to be a consistent truncation, there will be some UV cutoff scale $\Lambda$, which we can take to scale with $M_P$, at which corrections become important. Thus, we will take $x_{\text{min}}=1/(\Lambda\ell)$.

The prefactor $1/G\sim N^2$ in $F(x)$ makes it clear that the integrand grows like $N^2$ (like $C_T$) in the large $N$ limit, so that a saddle point approximation is appropriate. $F(x)$ has two saddle points, located at
\begin{align}x^s_\pm=\frac{ (2 d-3)\pi \pm \sqrt{\pi ^2 (2d-3)^2 -(\beta/\ell) ^2 (d-2)^2 (d-1) d}}{(\beta/\ell)  (d-1) d}.\end{align}
For $d>3$, there is a third saddle, located at $x=0$. These coincide with the extrema of $q^2$ as a function of $x$. This means that the minimum at $x_-^s$ is always in the first intervals of $R$, while the maximum at $x_+^s$ is not in $R$ (this corresponds to the minimum of $q^2$, which is necessarily negative). The maxima and minima disappear precisely at  $\beta=\beta_c$, when the two intervals of $R$ join into one; this means that for low temperatures there are no saddles (other than the one at $x=0$) and $F(x)$ is monotonically decreasing.

In the first interval of $R$, the integral is dominated by the contributions near $x=0$; the saddle at $x_-^s$ is always subdominant. This means that the integral is controlled by the behavior at $x_{\text{min}}$. In the second interval, the integrand is a decreasing function of $x$, so that the integral is controlled by its value around $x_+$. If $F(x_+)>0$, this contribution will grow as $e^{F(x_+)}$ in the large $N$ limit, and therefore will dominate $\mathcal{Z}(\beta)$; if $F(x_+)<0$, the saddle will be again subdominant and the dominant contribution will come from $x\approx0$. The transition point $F(x_+)=0$ happens precisely at the Hawking-Page temperature $\beta_{HP}$, eq. \eq{bHP}. 

So, to sum up, for $\beta>\beta_{HP}$, the partition function is dominated by the region around $x\approx x_\text{min}$, which is the maximum in the integration region; by contrast, for $\beta>\beta_{HP}$, $\mathcal{Z}$ will be dominated by the contribution around $x=x+$, where $F$ again attains its maximum over $R$. All that is left is to evaluate the contributions at $x=x_{\text{min}},x_+$:\begin{itemize}
\item Since $x=x_{\text{min}}$ is not a maximum, the leading term in the Taylor expansion is linear in $x$, so that a change of variables $u^2=(x-x_{\text{min}})$ makes $F$ quadratic. Applying again the saddle point formula, we obtain
\begin{align}\mathcal{Z}(\beta)&\approx \frac{\kappa_{d+1}}{g\ell}\,e^{F(x_{\text{min}})}\, \frac{d-2}{\sqrt{(\beta/\ell) (d-1) \left((\beta/\ell) \left(d x_{\text{min}}^2+d-2\right)-4 \pi  x_{\text{min}}\right)}}.\label{xzero}\end{align}
\item Again $x=x_+$ is not a maximum, so doing the same as above, we get \eq{xzero} again, with $x_+$ instead of $x_{\text{min}}$.

Taking logarithms and dropping the subleading contribution coming from the last factor of \eq{xzero}, we recover equation \eq{key} of the main text.

\end{itemize}

\bibliographystyle{jhep}
\bibliography{gcbound}

\end{document}